\documentclass{sig-alternate}
\pdfoutput=1
\RequirePackage{voila}
\RequirePackage[T1]{fontenc}
\RequirePackage{hyperref}

\belowdisplayskip \abovedisplayskip
\abovedisplayshortskip \abovedisplayskip
\belowdisplayshortskip \abovedisplayskip
\begin{document}

\title{Learning Joint Query Interpretation and Response Ranking}
\author{Uma Sawant\thanks{\protect\path{uma@cse.iitb.ac.in}} \\ IIT Bombay
 \and Soumen Chakrabarti \\ IIT Bombay}
\maketitle

\begin{abstract}
Thanks to information extraction and semantic Web efforts,
search on unstructured text is increasingly refined using semantic
annotations and structured knowledge bases.  However, most users
cannot become familiar with the schema of knowledge bases and ask
structured queries.  Interpreting free-format queries into a more
structured representation is of much current interest.  The dominant
paradigm is to segment or partition query tokens by purpose
(references to types, entities, attribute names, attribute values,
relations) and then launch the interpreted query on 
structured knowledge bases.  Given that structured knowledge
extraction is never complete, here we use a data representation
that retains the unstructured text corpus, along with structured
annotations (mentions of entities and relationships) on it.  
We propose two new, natural formulations for joint query
interpretation and response ranking that exploit bidirectional flow of
information between the knowledge base and the corpus.
One, inspired by
probabilistic language models, computes expected response scores
over the uncertainties of query interpretation.  The other is based on
max-margin discriminative learning, with latent variables representing
those uncertainties.  In the context of typed entity search, both
formulations bridge a considerable part of the accuracy gap between
a generic query that does not constrain the type at all, and
the upper bound where the ``perfect'' target entity type of each query is 
provided by humans.  Our formulations are also superior to a two-stage 
approach of first choosing a target type using recent query type 
prediction techniques, and then launching a type-restricted 
entity search query.
\end{abstract}


\section{Introduction}
\label{sec:Intro}

Web information representation is getting more sophisticated, thanks
to information extraction and semantic Web efforts.  Much structured
and semistructured data now supplements unstructured, free-format
textual pages.  In verticals such as e-commerce, the structured data 
can be accessed through forms and faceted search.  However, a
large number of free-format queries remain outside the scope of
verticals.  As we shall review in Section~\ref{sec:Rel}, there is much
recent research on analyzing and annotating them.

Here we focus on a specific kind of entity search query: Some words
(called \emph{selectors}) in the query are meant to occur literally in
a response document (as in traditional text search), but other words
\emph{hint} at the type of entity sought by the query.  Unlike prior
work on translating well-formed sentences or questions to structured
queries using deep NLP, we are interested in handling ``telegraphic''
queries that are typically sent to search engines.  Each response
entity must be a member of the hinted type.

Note that this problem is quite different from finding answers to
well-formed natural language questions (e.g., in Wolfram Alpha) 
from structured knowledge bases
(perhaps curated through information extraction).  Also observe that
we do not restrict ourselves to queries that seek entities by
attribute values or attributes of a given entity (both are valuable
query templates for e-commerce and have been researched).  In our
setup, some responses may only be collected from diverse, open-domain,
free-format text sources.  E.g., typical driving \emph{time} between
Paris and Nice (the target type is time duration), or
\emph{cricketers} who scored centuries at Lords (the target type is
cricketers).

The target type (or a more general supertype, such as
\emph{sportsperson} in place of \emph{cricketer}) may be instantiated
in a \emph{catalog}, but the typical user has no knowledge of the catalog or
its schema.  Large catalogs like Wikipedia or Freebase evolve
``organically''.  They are not designed by linguists, and they are not
minimal or canonical in any sense.  Types have overlaps and
redundancies.  The query interpreter should take advantage of
specialized types whenever available, but otherwise gracefully
back off to broader types.

\begin{figure}[th]
\centering\includegraphics{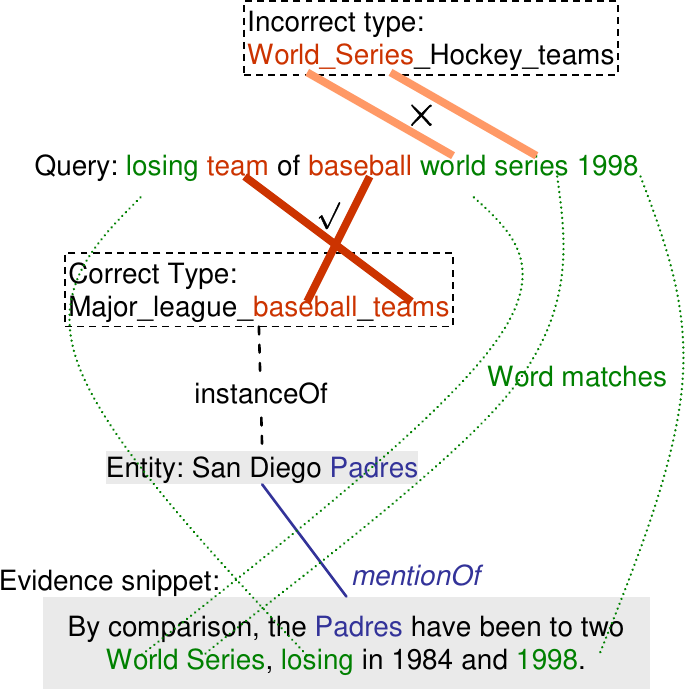}
  \caption{A partition of query words into hints and selectors, some
    partially matching types with member entities, and corpus snippets
    constitute a collective, joint query interpretation and entity
    ranking problem.}
  \label{fig:Framework}
\end{figure}

Figure~\ref{fig:Framework} shows a query that has at least two plausible
hint word sets: $\{\text{team}, \text{baseball}\}$ (correct) and
$\{\text{world}, \text{series}\}$ (incorrect).  Hint words partially
match descriptions of types in a catalog, which lead to member
entities.  Potential response entities are mentioned in document
snippets (one shown), which in turn partially match selector words
(world, series, losing, 1998).  Given a limited number of 
types to choose from, a
human will find it trivial to pick the best.  However, a program
will find it very challenging to decide \emph{which subset} of query
words are type hints, and, even after that, to select the \emph{best
  type(s)} from a large type catalog.  This \emph{query
  interpretation} task is one part of our goal.

We posit that \emph{corpus statistics provide critical signals} for
query interpretation.  For example, we might benefit from knowing that
\texttt{San\_Diego\_Padres} rarely co-occurs with the word ``hockey'',
which can be known only from the corpus.  Query interpretation should
ideally be done \emph{jointly} with ranking entities from the corpus,
because it involves a \emph{delicate combinatorial balance} between
the hint-selector split, and the (rather noisy) signals from the quality
of matches between type descriptions and hint words, snippets and
other words, and mentions of entities in said snippets.

Although query typing has been investigated before
\cite{ValletZ2008TypeInQuery, BalogN2012TargetTypeEntityQueries}, to
the best of our knowledge this is the first work on combining type
interpretation with learning to rank \cite{Liu2009LearningToRank}.  In
Section~\ref{sec:Gen}, we present a natural, generative formulation
for the task using probabilistic language models.  In
Section~\ref{sec:Disc} we present a more flexible and powerful
max-margin discriminative approach \cite{Joachims2002ranksvm,
  BergeronZBB2008MIR}.

In Section~\ref{sec:Expt}, we report on experiments involving
709 queries, over 200,000 types, 1.5 million entities, and 380 million
evidence snippets collected from over 500
million Web pages.  The entity ranking accuracy of a reasonable query
interpreter will be between the ``lower bound'' of a generic system
that makes no effort to identify the target type (i.e., all catalog
entities are candidates), and the upper bound of an unrealistic
``perfect'' system that knows the target type by magic.  Our salient
experimental observations are:
\begin{itemize}
\item The generative language model approach improves entity ranking
  accuracy significantly beyond the lower bound wrt MAP, MRR and NDCG.
\item The discriminative approach is superior to generative; e.g., it
  bridges 43\% of the MAP gap between the lower and upper
  bounds.
\item In fact, if we discard the entity ranks output from our system,
  use it only as a target type predictor, and issue a query with the
  predicted type, entity ranking accuracy \emph{drops}.
\item Our discriminative approach beats a recent target type prediction
  algorithm by significant margins.
\item NLP-heavy techniques are not robust to telegraphic queries.
\end{itemize}
Our data and code will be made publicly available.


\section{Related work}
\label{sec:Rel}

Interpreting a free-format query into a structured form has been
explored extensively in the information retrieval (IR) and Web search
communities, with several recent dedicated
workshops\footnote{\protect\href{http://ciir.cs.umass.edu/sigir2010/qru/}
{ciir.cs.umass.edu/sigir2010/qru}
\protect\href{http://ciir.cs.umass.edu/sigir2011/qru/}{ciir.cs.umass.edu/sigir2011/qru}
\goodbreak \protect\href{http://strataconf.com/stratany2011/public/schedule/detail/21413}
{strataconf.com/stratany2011/public/schedule/detail/21413} \goodbreak
\protect\href{http://sysrun.haifa.il.ibm.com/hrl/smer2011/}{sysrun.haifa.il.ibm.com/hrl/smer2011} \goodbreak
\protect\href{http://km.aifb.kit.edu/ws/jiwes2012/}{km.aifb.kit.edu/ws/jiwes2012}}.
A preliminary but critical structuring step is to demarcate phrases
\cite{BenderskyCS2009QuerySegment} in free-format queries.
There is also a large literature on topic-independent intent discovery
\cite{Broder2002Taxonomy,JansenBS2008QueryIntent} as well as
topic-dependent facet \cite{PoundPT2011Facet} or template
\cite{AgarwalKC2010QueryTemplate} inference.

\begin{figure}[h]
\centering\includegraphics{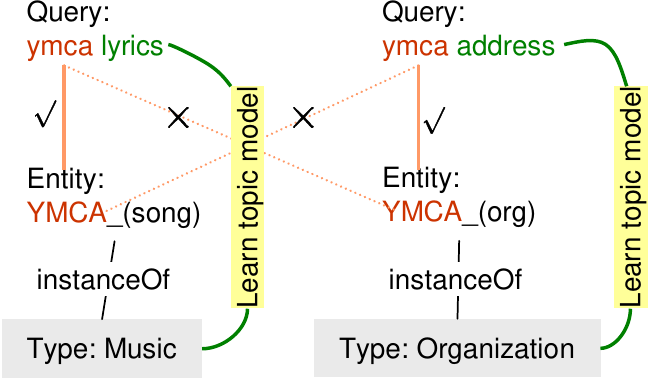}
  \caption{Disambiguating named entities in queries.}
  \label{fig:NERQ}
\end{figure}

The problem of \emph{disambiguating named entities} mentioned in
queries is superficially similar to ours, but is technically quite
different.  In Figure~\ref{fig:NERQ}, query word \emph{ymca} may refer
to different entities, but additional query word \emph{lyrics} hints
at type \emph{music}, whereas \emph{address} hints at type
\emph{organization}.  Note that the query text directly embeds a
mention of an \emph{entity}, not a type.  Disambiguating the entity
(usually) amounts to disambiguating the type---contrast
Figure~\ref{fig:NERQ} with Figure~\ref{fig:Framework}.  A given
mention usually refers to only a few entities.  In contrast,
misinterpreting the hint often pollutes the entity response
list beyond redemption.  Delaying a hard choice of 
the target type, or avoiding it entirely, is likely to help.

For entity disambiguation,
Guo \etal~\cite{GuoXCL2009nerq} proposed a probabilistic language
model through weak supervision that learns to associate, e.g., lyrics
with music and address with organization.  Pantel
\etal~\cite{PantelF2011JigsLures, PantelLG2012EntityTypeIntent} pushed
this farther by exploiting clicks and modeling intent.  Hu
\etal~\cite{HuWLSC2009QueryIntentWikipedia} addressed a similar
problem.  None gave a discriminative max-margin formulation,
or unified the framework with learning to rank.

Given that the database community uses SQL and XQuery as unambiguous,
structured representations of information needs, and that the NLP
community seeks to parse sentences to a well-defined meaning, there
also exists convergent database and NLP literature on interpreting
free-format (source) queries into a suitable target ``query
language''.  Naturally, much of this work seeks to identify types,
entities, attributes, and relations in queries.  Although the
theoretical problem is challenging
\cite{FaginKLRV2010UnderstandingQueries}, a common underlying theme is
that each token in the query may be an expression of schema elements,
entities, or relationships: this leads to a general assignment
problem, which is solved approximately using various techniques,
summarized below.

Sarkas \etal~\cite{SarkasPT2010QueryAnnotation} annotated e-commerce
queries using schema and data in a structured product catalog.  In the
context of Web-extracted knowledge bases such as YAGO
\cite{SuchanekKW2007YAGO}, Pound \etal~\cite{PoundIW2010WebSQL,
  PoundHIW2012InterpretQuery} set up a collective assignment problem
with a cost model that reflects syntactic similarity between query
fragments and their assigned concepts, as well as semantic coherence
between concepts~\cite{KulkarniSRC2009CSAW}.  Sarkas, Pound and
others, like us, handle ``telegraphic queries'' that may not be
well-formed sentences.  DEANNA \cite{YahyaBERTW2012DEANNA} solved the
collective assignment problem using an integer program.  It is capable
of parsing queries as complex as ``which director has won the Academy
Award for best director and is married to an actress that has won the
Academy Award for best actress?''  As might be expected, DEANNA is
rather sensitive to query syntax and often fails on telegraphic
queries.  All these systems interpret the query with the help of a
fairly clean, structured knowledge base.  
\cite{SarkasPT2010QueryAnnotation, PoundIW2010WebSQL,
  PoundHIW2012InterpretQuery, YahyaBERTW2012DEANNA} do
not give discriminative learning-to-rank algorithms
that jointly disambiguate the query and ranks responses.
IBM's Watson \cite{MurdockKWFFGZK2012TypeCoercion}
identifies candidate entities first, and then scores them
for compatibility with likely target types.

In this work, we do not assume that a knowledge base has been curated
ahead of time from a text corpus.  Instead we assume entities and
types have been annotated on spans of unstructured text.  Accordingly,
we step back from sophisticated target schemata, settling for three
basic relations (\emph{instanceOf}, \emph{subTypeOf}, and
\emph{mentionOf}, see Figure~\ref{fig:Framework}) that link a
structured entity catalog with an unstructured text corpus (such as
the Web).  On the other hand, we take the first step toward
integrating learning-to-rank \cite{Liu2009LearningToRank} techniques
with query interpretation.

Closest to our goal are those of Vallet and Zaragoza
\cite{ValletZ2008TypeInQuery} and Balog and Neumayer (B\&N)
\cite{BalogN2012TargetTypeEntityQueries}.  Vallet and Zaragoza first
collected a ranked list of entities by launching a query without any
type constraints.  Each entity belongs to a hierarchy of types.  They
accrued a score in favor of a type from every entity as a function of
its rank, and ranked types by decreasing total score.  B\&N investigated
two techniques.  In the first, descriptions of all entities $e$
belonging to each type $t$ were concatenated into a super document for
$t$, and turned into a language model.  In the second (similar in
spirit to Vallet and Zaragoza), the score of $t$ was calculated as a
weighted average of probabilities of entity description language
models generating the query, for $e\in t$.

These approaches \cite{BalogN2012TargetTypeEntityQueries, 
RavivCK2012MrfEntityRank} use long
entity descriptions, such as found on the Wikipedia page representing
an entity, but not a corpus where entity mentions are annotated.  The
corpus documents may well not be definitional, and yet remarkably
improves entity ranking accuracy, as we shall see.  None of
\cite{ValletZ2008TypeInQuery, BalogN2012TargetTypeEntityQueries,
  RavivCK2012MrfEntityRank} attempt a segmentation of query words by
purpose (target type vs.\ literal matches).


\section{Background and notation}
\label{sec:Setup}

\subsection{``Telegraphic'' queries}
\label{sec:Setup:Telegraph}

A ``telegraphic'' entity search query $q$ expresses an information need that is
satisfied by one or more \emph{entities}.   Query $q$ is a sequence of
$|q|$ \emph{words}.  The $j$th word of query $q$ is denoted $w_{q,j}$,
where $j=1,\ldots,|q|$, and subscript $q$ in $w_{q,j}$ is omitted if
clear from context.  We will interchangeably use $q$ (as a query
identifier) and $\vec q$ (to highlight that it is a sequence of
words).  Unlike full, well-formed, grammatical sentences or questions,
telegraphic queries resemble short Web search queries having no clear
subject-verb-object or other complex clausal structure.  Some examples
of natural telegraphic entity search queries and possible natural
language ``translations'' are shown in Figure~\ref{fig:Telegraph}.
$\mathcal{Q}$ denotes a set of queries.

\begin{figure}[tb]
  \begin{small}
\begin{tabular}{l|p{.43\hsize}|p{.43\hsize}}
\raggedright
Q1 &
Woodrow Wilson was president of which university? &
woodrow wilson \underline{president} \underline{university}
 \\ \hline
\raggedright
Q2 &
Which Chinese cities have many international companies? &
chinese \underline{city} many \mbox{international} \underline{companies}
 \\ \hline
\raggedright
Q3 &
What cathedral is in Claude Monet's paintings? &
\underline{cathedral} claude monet \underline{paintings}
 \\ \hline
\raggedright
Q4 &
Along the banks of what river is the Hermitage Museum located? &
hermitage \underline{museum} \underline{banks} of \underline{river} 
 \\ \hline
\raggedright
Q5 &
At what institute was Dolly cloned? & 
dolly \underline{clone} \underline{institute}
 \\ \hline
\raggedright
Q6 &
Who made the first airplane? & 
first \underline{airplane}
\underline{inventor}
\end{tabular}    
  \end{small}
  \caption{Natural language queries and typical telegraphic forms,
    with potential type description matches underlined.}
  \label{fig:Telegraph}
\end{figure}

\subsection{The entity and type catalog}
\label{sec:Setup:Catalog}

The \emph{catalog} $(\mathcal{T}, \mathcal{E}, \subseteq^+, \in^+)$,
is a directed acyclic graph of \emph{type} nodes $t\in\mathcal{T}$,
with edges representing the ``is-subtype-of'' transitive binary
relation $\subseteq^+$.  Each type $t$ is described by one or more
\emph{lemmas} (descriptive phrases) $L(t)$, e.g.,
\href{en.wikipedia.org/wiki/Category:Austrian_physicists}{Austrian
  physicists}.

Each entity $e$ in the catalog is also
represented by a node connected by ``is-instance-of'' edge(s) to one
or more \emph{most specific} type nodes, and transitively belongs to
all supertypes; this relation is represented as $\in^+$.  An entity
$e$ may be a \emph{candidate} for a query $q$.  The set of candidate
entities for query $q$ is called $\mathcal{E}_q \subseteq
\mathcal{E}$.  In training data, an entity $e$ may be labeled relevant
(denoted $e_+$) or irrelevant (denoted $e_-$) for~$q$.  
$\mathcal{E}_q$ is accordingly partitioned into 
$\mathcal{E}_q^+, \mathcal{E}_q^-$.

\subsection{Annotated corpus and snippets}
\label{sec:Setup:Snippets}

The corpus is a set of free-format text documents.  Each document is
modeled as a sequence of words.  Entity $e$ is \emph{mentioned} at
some places in an unstructured text corpus.  A ``mention'' is a token
span (e.g., \emph{Big Apple}) that gives evidence of reference to $e$
(e.g.,
\href{http://en.wikipedia.org/wiki/New_York_City}{New\_York\_City}).
The mention span, together with a suitable window of context words
around it, is called a \emph{snippet}.  The set of snippets mentioning
$e$ is called $\mathcal{S}_e$.  $c \in \mathcal{S}_e$ is one snippet
context \emph{supporting}~$e$.

In the Wikipedia corpus, most mentions are annotated manually as wiki
hyperlinks.  For Web text, statistical learning techniques
\cite{KulkarniSRC2009CSAW, Hoffart+2011RobustDisambiguation} are
used for high-quality annotations.  Here we assume mentions to be
correct and deterministic.  Extending our work to noisy mentions is
left for future work.


\section{Generative formulation}
\label{sec:Gen}

Given the success of generative techniques in corpus modeling
\cite{BleiNJ2003lda}, IR \cite{Zhai2008LanguageModels} and entity
ranking \cite{BalogAR2006ExpertSearch,
  BalogAdR2009LanguageModelExpert}, it is natural to propose a
generative language model approach to joint query interpretation and
response ranking.

As is common in generative language models, we will fix an entity $e$
and generate the query words, by taking the following steps:
\begin{enumerate}
\item Choose a type from $\{t : e \in^+ t\}$;
\item Describe that type using one or more query words, which will be
  called \emph{hint} words;
\item Collect snippets that mention $e$; and
\item Generate the remainder of the query by sampling words from these
  snippets.
\end{enumerate}
Our goal is to rank entities by probability given the query, by
taking the expectation over possible types and hints.

\subsection{Choosing a type given $e$}
\label{sec:Gen:PickType}

Given entity $e$, we first pick a type $t$ such that $e \in^+ t$, and
describe $t$ in the query (with the expectation that the system will
infer $t$, then instantiate it to $e$ as a response).  So the basic
question looks like: ``if the answer is Albert Einstein, what type
(among scientist, person, organism, etc.) is likely to be mentioned in
the query, \emph{before} we inspect the query?''  (\emph{After} we see
the query, our beliefs will change, e.g., depending on whether the
query asks ``\emph{who} discovered general relativity?'' vs.\ ``which
\emph{physicist} discovered general relativity?'')  So we need to
design the prior distribution $\Pr(t|e)$.

Recall that there may be hundreds of thousands of $t$s, and tens of
millions of $e$s, so fitting the prior for each $e$ separately is out
of the question.  On the other hand, the prior is just a mild guidance
mechanism to discourage obscure or low-recall types like ``Austrian
Physicists who died in 1972''.  Therefore, we propose the following
crude but efficient estimate.  From a query log with ground truth
(i.e., each query accompanied with a $t$ provided by a human),
accumulate a hit count $N_t$ for each type $t$.  At query time, given
a candidate $e$, we calculate
\begin{align}
\Pr(t|e) &=
  \begin{cases}
\displaystyle\frac{N_t + \gamma}{ \sum_{t': e\in^+ t'} (N_{t'} + \gamma)},
& e \in^+ t \\
0, & \text{otherwise}
  \end{cases},
\end{align}
where $\gamma\in(0,1)$ is a tuned constant.

\subsection{Query word switch variables}
\label{sec:Gen:Switch}

Suppose the query is the word sequence $(w_j, j = 1, \ldots, |q|)$.
For each position $j$, we posit a binary switch variable $z_j \in \{h,
s\}$.  Each $z_j$ will be generated iid from a Bernoulli distribution
with tuned parameter $\delta \in (0,1)$.  If $z_j = h$, then word
$w_j$ is intended as a \emph{hint} to the target type.  Otherwise
$w_j$ is a \emph{selector} sampled from snippets mentioning
entity~$e$.  The vector of switch variables is called $\vec z$.  

The number of possible partitions of query words into hints and
selectors is $2^{|q|}$.  By definition, telegraphic queries are short,
so $2^{|q|}$ is manageable.  One can also reduce this search space by
asserting additional constraints, without compromising quality in
practice.  E.g., we can restrict the type hint to a contiguous span
with at most three tokens.

Given $\vec q$ and a proposed partition $\vec z$, we define two helper
functions, overloading symbols $s$ and $h$:
\begin{align}
\text{Hint words of $q$:}&& h(\vec q,\vec z) &= \{w_{q,j} : z_j=h\}
\label{eq:HintDefn} \\
\text{Selector words of $q$:} && s(\vec q,\vec z) 
&= \{w_{q,j} : z_j=s\}. \label{eq:SelDefn}
\end{align}
With these definitions, in the exhaustive hint-selector partition
case, $\vec z$ is the result of $|q|$ Bernoulli trials with hint
probability $\delta\in(0,1)$ for each word, so we have
\begin{align}
\Pr(\vec z) &= \delta^{|h(\vec q,\vec z)|} (1-\delta)^{|s(\vec q,\vec z)|}.
\label{eq:PrZdefn}
\end{align}
$\delta$ is tuned using training data.

In this paper we will consider strict partitions of query
words between hints and selectors, but it is not difficult to
generalize to words that may be both hints and selectors.  Assuming
each query word has a purpose, the full space grows to $3^{|q|}$, but
assuming contiguity of the hint segment again reduces the space to
essentially $O(|q|)$.

\subsection{Type description language model}
\label{sec:Gen:TypeLM}

Globally across queries, the textual description of
each type $t$ induces a language model.  We can define the exact form of the
model in any number of ways, but, to keep implementations efficient,
we will make the commonly used assumption that hint words are
conditionally independent of each other given the type.  Each type $t$ is described
by one or more \emph{lemmas} (descriptive phrases) $L(t)$, e.g.,
\href{en.wikipedia.org/wiki/Category:Austrian_physicists}{Austrian
  physicists}.  Because lemmas are very short, words are rarely
repeated, so we can use the multivariate Bernoulli
\cite{McCallumN1998event} distribution derived from lemma~$\ell$:
\begin{align}
&&  \widehat{\Pr}(w|\ell) &= 
  \begin{cases}
    1, & \text{if $w$ appears in $\ell$}, \\
    0, & \text{otherwise}
  \end{cases}  \label{eq:TlmRough}
\end{align}
Following usual smoothing policies \cite{Zhai2008LanguageModels}, we
interpolate the smoothed distribution above with a background language
model created from all types:
\begin{align}
\widehat{\Pr}(w|\mathcal{T}) = 
\frac{\sum_{t \in \mathcal{T}}
\llbracket \text{$w$ appears in $\ell$ ; $\ell \in L(t)$} \rrbracket}
{ |\mathcal{T}|};  \label{eq:TlmBackground}
\end{align}
in words, the fraction of all types that contain $w$.  We splice
together \eqref{eq:TlmRough} and \eqref{eq:TlmBackground} using
parameter $\beta \in(0,1)$:
\begin{align}
\Pr(w|\ell) = (1-\beta) \widehat{\Pr}(w|\ell) +
\beta \widehat{\Pr}(w|\mathcal{T}).
\end{align}
The probability of generating exactly the hint words in the query is
\begin{align}
\Pr(h(\vec q,\vec z)|\ell) &= \prod_{w\in h(\vec q,\vec z)} \Pr(w|\ell)
\prod_{w\not\in h(\vec q,\vec z)} (1-\Pr(w|\ell)), \label{eq:PrHint}
\end{align}
where $w$ ranges over the entire vocabulary of type descriptions.  In
case of multiple lemmas describing a type,
\begin{align}
\Pr(\cdot |t) &= \max_{\ell \in L(t)} \Pr(\cdot |\ell);
\label{eq:PrHintBestLemma}
\end{align}
i.e., use the most favorable lemma.
All fitted parameters in the distribution $\Pr(w|\ell)$ are
collectively called $\varphi$.

\subsection{Entity snippet language model}
\label{sec:Gen:SnippetLM}

The selector part of the query, $s(\vec q,\vec z)$, 
is generated from a language model derived from
$\mathcal{S}_e$, the set of snippets that mention candidate
entity~$e$.  For simplicity we use the same kind of smoothed
multivariate Bernoulli distribution to build the language model as we
did for the type descriptions.  Note that words that appear in
snippets but not in the query are of no concern in a language model
that seeks to generate the query from distributions associated with
the snippets.  Suppose $\corpusCount(e)$ is the number of mentions of
$e$ in the corpus $\mathcal{C}$, and $\corpusCount(e,w)$ be the number of mentions
of $e$ where $w$ also occurs within a specified snippet window width.
The unsmoothed probability of
generating a query word $w$ from the snippets of $e$ is
\begin{align}
 \widehat{\Pr}(w|e) &= \frac{\corpusCount(e,w)}{\corpusCount(e)}
= \frac{\left|\{s\in \mathcal{S}_e: w \in s\}\right|}{\corpusCount(e)}.
\end{align}
As before, we will smooth the above estimate using an corpus-level,
entity-independent background word distribution estimate:
\begin{align}
  \widehat{\Pr}(w|\mathcal{C}) &=
  \frac{1}{|\mathcal{C}|}
       (\text{number of documents containing $w$}).
\end{align}
And now we use the interpolation
\begin{align}
\Pr(w|e) &= (1-\alpha) \widehat{\Pr}(w|e)
+ \alpha \widehat{\Pr}(w|\mathcal{C}),
\end{align}
where $\alpha\in(0,1)$ is a suitable smoothing parameter.  The fitted
parameters of the $\Pr(w|e)$ distribution are collectively
called~$\theta$.  Similar to \eqref{eq:PrHint}, the selector part of
the query is generated with probability
\begin{align}
  \Pr(s(\vec q, \vec z)|e) = 
\prod_{w\in s(\vec q, \vec z)} \Pr(w|e)
\prod_{w\not\in s(\vec q, \vec z)}(1-\Pr(w|e)),
\label{eq:PrSelFromEntSnips}
\end{align}
except here $w$ ranges over all query words.

\begin{figure}[h]
\centering\includegraphics{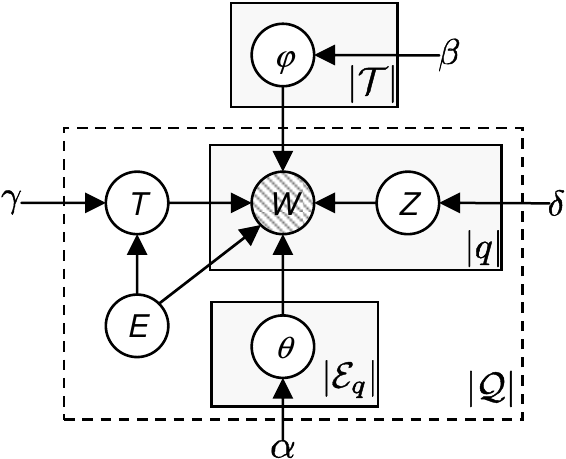}
\caption{Plate diagram for generating a query $q$ from a
  candidate entity $e$.  Only $(w_{q,j}:j=1,\ldots,|q|)$ are observed
  variables.  $\varphi$ represents the type description language model
  and $\theta$ represents the entity mention snippets language model.
  $(z_{q,j}:j=1,\ldots,|q|)$ are the hidden switch variables.  $T$ is
  the hidden type variable.} \label{fig:Plate}
\end{figure}

\subsection{Putting the pieces together}
\label{sec:Gen:Glue}

A plate diagram for the process generating a query $\vec q$ is shown
in Figure~\ref{fig:Plate}.  Vertices are marked with random variables
$E, T, Z, W$ whose instantiations are specific values $e, t, \vec z,
w\in q$.

The hidden variables of interest are the binary $Z\in\{h,s\}$, for
selecting between type hint ($h$) and selector ($s$) words; and $T$,
the type of one query.  Each query picks one hidden value $t$,
and a vector of $|q|$ size for $Z$, denoted $\vec z$.  The only
observed variables are the $|q|$ query words $(w_j: j=1,\ldots,|q|)$.
Also, $\alpha,\beta,\gamma,\delta$ are hyper-parameters tuned globally
across queries.

In the end we are interested in $\arg\max_e \Pr(e|\vec q)$, where
\begin{align}
\Pr(e|\vec q) &\propto \Pr(e, \vec q)
= \Pr(e) \Pr(\vec q|e) = 
 \Pr(e) \sum_{t, \vec z} \Pr(\vec q, t, \vec z|e) \notag\\
&= \Pr(e) \sum_{t,\vec z} \Pr(t|e) \underline{\Pr(\vec z|e,t)} 
\Pr(\vec q|e, t, \vec z) \label{eq:PrZgivenET} \\
&\approx
\Pr(e) \sum_{t,\vec z} \Pr(t|e) \underline{\Pr(\vec z)}
\Pr(\vec q|e, t, \vec z) \label{eq:PrZ} \\
&= \Pr(e) \sum_{t, \vec z} \Pr(t|e) 
\underbrace{\Pr(\vec z)}_{\eqref{eq:PrZdefn}}
\underbrace{\Pr(h(\vec q,\vec z)|t)}_{\eqref{eq:PrHintBestLemma}}
\underbrace{\Pr(s(\vec q,\vec z)|e)}_{\eqref{eq:PrSelFromEntSnips}}. 
\notag
\end{align}
To get from \eqref{eq:PrZgivenET} to \eqref{eq:PrZ} we make the
simplifying assumption that the density of hint words in queries is
independent of the candidate entity and type.  As mentioned before,
adding over $t, \vec z$ is feasible for telegraphic queries because
they are short.  The prior $\Pr(e)$ may be uninformative (i.e.,
uniform), or set proportional to $|\mathcal{S}_e|$ 
\cite{MacdonaldO2011RankingAggregates}, or use shrunk
estimates from answer types in the past.  We use
$\Pr(e)=|\mathcal{S}_e|/\sum_{e'}|\mathcal{S}_{e'}|$.

If we allow a query word to represent both a type hint and a selector,
the clean separation after \eqref{eq:PrZ} no longer works, but it is
possible to extend the framework using a soft-OR expression.  We omit details
owing to space constraints.

\subsection{Explaining a top-ranking entity}
\label{sec:Gen:Explain}

In standard text search, top-ranking URLs are accompanied by a summary
with matching query words highlighted.  In our system, top-ranking
entities need to be justified by explaining to the user how the query
was interpreted.  Specifically, we need to show the user the inferred
type, and the inferred purpose (hint or selector) of each query word.
\begin{align}
\Pr(t, \vec z|e,\vec q) &\propto \Pr(e,t,\vec q,\vec z) \notag \\
&= \Pr(e) \Pr(t|e) \underline{\Pr(\vec z|e,t)} 
\Pr(\vec q|e, t, \vec z) \notag \\
&\approx \Pr(e) \Pr(t|e) \underline{\Pr(\vec z)}
\Pr(\vec q|e, t, \vec z) 
\end{align}
approximating $\Pr(\vec z|e,t)\approx\Pr(\vec z)$ as before.  Now we
can report $\arg\max_{t, \vec z} \Pr(t,\vec z|e, \vec q)$ as the
explanation for~$e$.  It is also possible to report marginals such as
$\Pr(t|e,\vec q)$ or $\Pr(z_j|e,\vec q)$ this way.

\subsection{Potential pitfalls}
\label{sec:Gen:Bad}

As often happens, a generative formulation starts out feeling natural,
but is soon mired in a number of questionable assumptions and tuned
hyper parameters.  In recent times, this story has played out in many
problems, such as information extraction \cite{Sarawagi2008IE} and
learning to rank \cite{Liu2009LearningToRank}, where generative
language models were proposed earlier, but the latest algorithms are
all discriminatively trained.  The above formulation has several
potential shortcomings:
\begin{itemize}
\item 
The modeling of $\Pr(t|e)$ is necessarily a compromise.
\item 
$\Pr(z_j)$ is assumed to be independent of $q$ and $e$, and iid. These
  assumptions may not be the best.
\item 
In the interest of computational feasibility, the language models for
both types and snippets are simplistic.
Phrase and exact matches are difficult to capture.
\item 
Hyper parameters $\alpha,\beta,\gamma,\delta$ can only be tuned by
sweeping ranges; no effective learning technique is obvious.
\item 
As often happens with complex generative models, the scales of
probabilities being multiplied \eqref{eq:PrZ} are diverse and hard to
balance.
\end{itemize}


\section{Discriminative formulation}
\label{sec:Disc}

Instead of designing conditional distributions as in
Section~\ref{sec:Gen}, here we will design feature functions, and
learn weights corresponding to them by using relevant and (samples of)
irrelevant entity sets $\mathcal{E}_q^+, \mathcal{E}_q^-$ associated
with each query $q$, as is standard in learning to rank
\cite{Liu2009LearningToRank}.
The benefit is that it is much safer to incrementally add highly 
informative but strongly correlated features (such as exact phrase match, 
match with and without stemming, etc.) to discriminative formulations.

Standard notation used in
structured max-margin learning uses $\phi(x,y) \in \R^d$ as the
feature map, where $x$ is an observation and $y$ is the label to be
predicted.  A \emph{model} $\lambda \in\R^d$ is fitted so that
$\lambda\cdot\phi(x,y_\text{correct}) > \lambda
\cdot\phi(x,y_\text{incorrect})$.  Once $\lambda$ gets fixed via
training, given a new text instance $x_\text{test}$, \emph{inference}
is the process of finding $\arg\max_y
\lambda\cdot\phi(x_\text{test},y)$.

In our case, we use the notation $\phi(q,e,t,\vec z)$ for the feature
map.  $q$ gives us access to the sequence of words in the query, and
is the analog of $x$ above.  $e$ gives us access to the snippets
$\mathcal{S}_e$ that support~$e$, and is the analog of $y$ above.  $t$
and $\vec z$ are \emph{latent variable}
\cite{YuJ2009LatentVariableSVM} inputs to the feature map whose role
will be explained shortly.

Guided by the generative formulation in Section~\ref{sec:Gen}, we
partition the feature vector as follows:
\begin{align}
\phi(q,e,t,\vec z) = 
\bigl(\phi_1(q,e), \phi_2(t,e), \phi_3(q,\vec z,t), \phi_4(q,\vec z,e) 
\bigr),  
\end{align}
where
\begin{itemize}
\item $\phi_1(q,e)$ models the prior for $e$.
\item $\phi_2(t,e)$ models the prior $\Pr(t|e)$.
\item $\phi_3(q,\vec z,t)$ models the compatibility between the type
  hint part of query words and the proposed type $t$.
\item $\phi_4(q,\vec z, e)$ models the compatibility between the
  selector part of query words and $\mathcal{S}_e$.
\end{itemize}

\subsection{Features $\phi_1$ modeling entity prior}

In Section~\ref{sec:Gen:Glue} we used
$\Pr(e)=|\mathcal{S}_e|/\sum_{e'}|\mathcal{S}_{e'}|$ as a prior
probability for~$e$.  It is natural to make this one element in
$\phi_1$.  But the discriminative setup allows us to introduce other
powerful features.

$|\mathcal{S}_e|$ does not distinguish between snippets that match the
query well vs.\ poorly.  Let $\IDF(w)$ be the inverse document
frequency \cite{BaezaYatesR1999MIR} of query word $w$, and $\IDF(q) =
\sum_{w\in q}\IDF(w)$.  $c \cap q$ is the set of query words found
in snippet $c$, with total $\IDF(c \cap q) = \sum_{w \in c \cap q}
\IDF(w)$.  Then the match-quality-weighted snippet support for $e$ is
characterized as
\begin{align}
\phi_1(q,e)[\cdot] &=
\frac{1}{2^{|q|} \IDF(q)}
\sum_{c \in \mathcal{S}_e} \IDF(c \cap q),
\end{align}
where $2^{|q|} \IDF(q)$ normalizes the feature across diverse queries.

Another feature in $\phi_1$ relates to negative evidence.  If
there are other words present, a query that directly mentions an
entity is hardly ever answered correctly by that entity;
\href{http://en.wikipedia.org/wiki/Tom_Cruise}{Tom\_Cruise} could not
be the answer for the query \texttt{tom cruise wife}.  Another (0/1)
element in $\phi_1$ is whether a description (``lemma'') of $e$ is
contained in the query.  In our experiments, the model element in
$\lambda$ corresponding to this feature turns out a negative
number, as expected.

\subsection{Features $\phi_2$ modeling type prior}

We have already proposed one way to estimate $\Pr(t|e)$ in
Section~\ref{sec:Gen:PickType}.  This estimate a natural element in
$\phi_2$.  We can also help the learner use the generality or
specificity of types, measured as this feature: ${\left|\{e: e
  \in^+ t \} \right|}/{|\mathcal{E}|}$.  In our experiments, the
element of $\lambda$ corresponding to this feature also got negative
values, indicating preference of specific types over generic ones.
This corroborates earlier observation regarding the depth of desired
types in a hierarchy~\cite{BalogN2012TargetTypeEntityQueries}.

\subsection{Hint-type compatibility features $\phi_3$}
\label{sec:Disc:HintType}

Given the input parameters of $\phi_3(\vec q,\vec z,t)$, we compute
the hint word subsequence $h(\vec q,\vec z)$ as in
\eqref{eq:HintDefn}.  Now we can define any number of features between
these hint words and the given type $t$, which has lemma set $L(t)$.
\begin{itemize}
\item
A standard feature borrowed from \eqref{eq:PrHintBestLemma} is
$\Pr(h(\vec q,\vec z)|t)$.
\item 
Unlike in the generative formulation, we can add synthetic features.
E.g., a feature that has value 1 if $\ell$ matches the subsequence
$h(\vec q,\vec z)$ \emph{exactly}.
\item
In Section~\ref{sec:Gen}, the size of $h(\vec q,\vec z)$ was drawn
from a binomial distribution controlled by hyper parameter $\delta$.
To model more general distributions, we use binary features of the
form
\begin{align*}
  \begin{cases}
    1, & |h(\vec q,\vec z)| < k \\
    0, & \text{otherwise}
  \end{cases}
\end{align*}
for $k = 1,\ldots$, to capture the belief that smaller number of hint
words is preferable.
\end{itemize}

\subsection{Selector-snippets compatibility features $\phi_4$}
\label{sec:Disc:SelSnip}

Now consider $q$ and its selectors $s(\vec q,\vec z) \subseteq q$ as
word sets (no duplicates), and the snippets $\mathcal{S}_e$ supporting
candidate entity~$e$.  $\phi_4(q,\vec z,e)$ will include feature/s
that express the extent of match or compatibility between the selector
words and the snippets.  We need to characterize and then combine two
kinds of signals here:
\begin{itemize}
\item The rarity (hence, informativeness) of a subset of $s(\vec
  q,\vec z)$ that match in snippets, and
\item The number of supporting snippets
  \cite{MacdonaldO2011RankingAggregates} that match a given word set.
\end{itemize}
(A third kind of signal, proximity
\cite{PetkovaBC2007NamedEntityProximity, TaoZ2007proximity,
  SvoreKK2010Proximity}, is favored indirectly, because snippets have
limited width.  A more refined treatment of proximity is left for
future work.)

A snippet $c \in \mathcal{S}_e$, interpreted as a subset of query
words $q$, \emph{covers} $s(\vec q,\vec z)$ if $c \supseteq s(\vec
q,\vec z)$.  Otherwise $c \subset s(\vec q, \vec z)$.  Recall every
snippet $c$ has an $\IDF(c) = \sum_{w \in c \cap q} \IDF(w)$.  We
propose two features:
\begin{align}
&& \, & \frac{1}{2^{|q|}\,\IDF(q)} \sum_{c \supseteq s(\vec q,\vec z)} \IDF(s(\vec q,\vec z))
\notag \\
&& &= \frac{\IDF(s(\vec q,\vec z))\,\left|
\{c: c \supseteq s(\vec q,\vec z) \} \right|}{2^{|q|}\,\IDF(q)}
\label{eq:CoverSupport} \\
\text{and} && \, & \frac{1}{2^{|q|}\,\IDF(q)} \sum_{c \subset s(\vec q,\vec z)} \IDF(c).
\label{eq:SubsetSupport}
\end{align}
We found the separation above to be superior to collapsing covering
and non-covering snippets into one sum.  Another useful feature was
the fraction of snippets $c$ such that $c = q$
(exactly matching all query words).


\subsection{Inference and training}

With a wrong choice of hint-selector partition $\vec z$, or a wrong
choice of type $t$, even a highly relevant response $e$ could score
very poorly.  Therefore, any reasonable scoring scheme should evaluate
$e$ under the \emph{best} choice of $t, \vec z$.  I.e., the score of
$e$ should be
\begin{align}
  \max_{t: e\in^+ t, \vec z} \lambda \cdot \phi(q, e, t, \vec z).
\label{eq:DiscInf}
\end{align}
(Note that $t$ ranges over only those types to which $e$ belongs.)  In
learning to rank \cite{Liu2009LearningToRank}, three training
paradigms are commonly used: itemwise, pairwise and listwise.  Because
of the added complexity from the latent variables $t, \vec z$, here we
discuss itemwise and pairwise training.  Listwise training is left for
future work.

In itemwise training, each response entity $e$ is one item, which can
be good (relevant, denoted $e_+$) or bad (irrelevant, denoted $e_-$).
Following standard max-margin methodology, we want
\begin{align}
\forall q, e_+: && \max_{t, \vec z} \lambda \cdot \phi(q, e_+, t, \vec z)
  & \ge 1 - \xi_{q,e_+}, \; \text{and} \label{eq:PosEnt} \\
\forall q, e_-: && \max_{t, \vec z} \lambda \cdot \phi(q, e_-, t, \vec z)
  & \le 1 + \xi_{q,e_-}, \label{eq:NegEnt}
\end{align}
where $\xi_{q,e_+}, \xi_{q,e_-} \ge 0$ are the usual SVM-style slack
variables.  Constraint \eqref{eq:NegEnt} is easy to handle by breaking
it up into the conjunct:
\begin{align}
\forall q, e_-, \forall t, \forall \vec z: && 
\lambda \cdot \phi(q, e, t, \vec z) & \le 1 + \xi_{q,e_-}.
\end{align}
However, \eqref{eq:PosEnt} is a \emph{disjunctive} constraint, as also
arises in multiple instance classification or ranking
\cite{BergeronZBB2008MIR}.  A common way of dealing with this is to
modify constraint \eqref{eq:PosEnt} into
\begin{align}
\forall q,e_+: && \sum_{t, \vec z} u(q, e_+, t,\vec z)
\lambda \cdot \phi(q, e_+, t, \vec z) & \ge 1 - \xi_{q,e_+}
\label{eq:Mic}
\end{align}
where $u(q,e,t,\vec z) \in \{0, 1\}$ and 
\begin{align*}
\forall q,e_+: && \sum_{t, \vec z} u(q, e_+, t,\vec z) &= 1.
\end{align*}
This is an integer program, so the next step is to relax the new
variables to $0 \le u(q,e,t,\vec z) \le 1$ (i.e., the $(t,\vec
z)$-simplex).  Unfortunately, owing to the introduction of new
variables $u(\cdots)$ and multiplication with old variables $\lambda$,
the optimization is no longer convex.

Bergeron \etal~\cite{BergeronZBB2008MIR} propose an alternating
optimization: holding one of $u$ and $\lambda$ fixed, optimize the
other, and repeat (there are no theoretical guarantees).  Note
that if $\lambda$ is fixed, the optimization of $u$ is a simple linear
program.  If $u$ is fixed, the optimization of $\lambda$ is comparable
to training a standard SVM.  The objective would then take the form
\begin{align}
  \tfrac{1}{2} \|\lambda\|^2 + \frac{C}{|\mathcal{Q}|}
\sum_{q \in \mathcal{Q}}
\frac{\sum_{e_+\in \mathcal{E}_q^+} \xi_{q,e_+} + 
\sum_{e_- \in \mathcal{E}_q^-} \xi_{q,e_-} }
{|\mathcal{E}_q^+| + |\mathcal{E}_q^-|}  \label{eq:PreAnneal}
\end{align}
Here $C>0$ is the usual SVM parameter trading off training loss against
model complexity.  Note that $u$ does not appear in the objective.

In our application, $\phi(q, e, t, \vec z) \ge \vec0$.  Suppose
$\lambda \ge \vec 0$ in some iteration (which easily happens in our
application).  In that case, to satisfy constraint \eqref{eq:Mic}, it
suffices to set only one element in $u$ to 1, corresponding to
$\arg\max_{t,\vec z} \lambda \cdot \phi(q, e, t, \vec z)$, and the
rest to 0s.  This severely restricts the search space over $u,
\lambda$ in subsequent iterations.

To mitigate this problem, we propose the following annealing protocol.
The $u$ distribution collapse reduces entropy suddenly.  The remedy is
to subtract from the objective (to be minimized) a term related to the
entropy of the $u$ distribution:
\begin{align}
\eqref{eq:PreAnneal}
+ D \sum_{q,e_+} \sum_{t, \vec z} u(q, e_+, t,\vec z) 
\log u(q, e_+, t,\vec z).
\label{eq:Anneal}
\end{align}
Here $D\ge 0$ is a temperature parameter that is gradually reduced in
powers of 10 toward zero with the alternative iterations optimizing $u$ and
$\lambda$.  Note that the objective \eqref{eq:Anneal} 
is convex in $u$, $\lambda$ and
$\xi_*$.  Moreover, with either $u$ or $\lambda$ fixed, all
constraints are linear inequalities.

\begin{figure}[h]
\begin{boxedminipage}{\hsize}
\begin{algorithmic}[1]
\STATE initialize $u$ to random values on the simplex
\STATE initialize $D$ to some positive value
\WHILE{not reached local optimum}
  \STATE fix $u$ and solve quadratic program to get next $\lambda$
  \STATE reduce $D$ geometrically
  \STATE fix $\lambda$ and solve convex program for next $u$
\ENDWHILE
\end{algorithmic}
\end{boxedminipage}
  \caption{Pseudocode for discriminative training.}
  \label{fig:DiscTrain}
\end{figure}

Very little changes if we extend from itemwise to pairwise training,
except the optimization gets slower, because of the sheer number of
pair constraints of the form:
\begin{align}
\forall q, e_+, e_-: &&
\max_{t, \vec z} \lambda \cdot \phi(q, e_+, t, \vec z) &-
\max_{t, \vec z} \lambda \cdot \phi(q, e_-, t, \vec z) \notag \\
&& &\ge 1 - \xi_{q, e_+, e_-}.  \label{eq:PairConstraint}
\end{align}
The itemwise objective in \eqref{eq:PreAnneal} changes to the pairwise
objectice
\begin{align}
  \tfrac{1}{2} \|\lambda\|^2 + \frac{C}{|\mathcal{Q}|}
\sum_{q \in \mathcal{Q}}
\frac{1}{|\mathcal{E}_q^+|\,|\mathcal{E}_q^-|}
\sum_{e_+ \in \mathcal{E}_q^+, e^- \in \mathcal{E}_q^-} \xi_{q, e_+, e_-}.
\end{align}
$u$-like variables can be used to convert this to an alternating
optimization as before; details are omitted.

\subsection{Implementation details} 

\subsubsection{Reducing computational requirements}

The space of $(q, e, t, \vec z)$ and especially their discriminative
constraints can become prohibitively large.  To keep RAM and CPU needs
practical, we used the following policies; our experimental results
are insensitive to them.
\begin{itemize}
\item
We sampled down bad (irrelevant) entities $e_-$ that were allowed to
generate constraint \eqref{eq:PairConstraint}.
\item
For empty $h(\vec q,\vec z) = \varnothing$,
$\phi_3(q,\vec z,t)$ provides no signal.  In such cases, we allow 
$t$ to take only one value: the most generic type \texttt{Entity}.

\end{itemize}

\subsubsection{Explaining a top-ranking entity}

This is even simpler in the discriminative setting than in the
generative setting; we can simply use \eqref{eq:DiscInf} to report
$\arg\max_{t,\vec z} \lambda \cdot \phi(q, e, t, \vec z)$.

\subsubsection{Implementing a target type predictor}
\label{sec:Disc:TypePredict}

Extending the above scheme, each entity $e$ scores each candidate
types $t$ as $\score(t|e) = \max_{\vec z} \lambda \cdot \phi(\cdot, e,
t, \vec z)$.  This induces a ranking over types for each entity.  We
can choose the overall type predicted by the query as the one whose
sum of ranks among the top-$k$ entities is smallest.  An apparently
crude approximation would be to predict the best type for the single
top-ranked entity.  But $k>1$ can stabilize the predicted type,
in case the top entity is incorrect.  
(We may want to predict a single type as a
feedback to the user, or to compare with other type prediction
systems, but, as we shall see, not for the best quality of entity
ranking, which is best done collectively.)


\section{Experiments}
\label{sec:Expt}

\subsection{Testbed}

\subsubsection{Catalog and annotated corpus}

Our type and entity catalog was YAGO \cite{SuchanekKW2007YAGO}, with
about 200,000 types and 1.9 million entities.  An
annotator trained on mentions of these entities in Wikipedia was run
over a Web corpus from a commercial search engine, having 500 million
spam-free Web pages.  This resulted in about 8 billion entity
annotations, average 16 annotations per page.  These were then indexed
\cite{ChakrabartiKBRS2012compressed}.

\subsubsection{Type constrained entity search}
\label{sec:Expt:QueryProcessor}

The index supports semistructured queries of the following form:
\begin{itemize}
\item an answer type $t$ from among the 200,000 YAGO types,
\item a bag of words and phrases in a IDF-WAND (weak-and) operator
  \cite{BroderCHSZ2003wand}, and
\item a snippet window width.
\end{itemize}
A DAAT \cite{BroderCHSZ2003wand} query processor returns a stream of
snippets at most as wide as the given window width limit, that contain
a mention of some entity $e \in^+ t$ and satisfies the WAND predicate.
In case of phrases in the query, the WAND threshold is computed by
adding the IDF of constituent words.  

Our query processor is implemented using MG4J \cite{BoVTREC2005} in
Java, with no index caching.  Basic keyword WAND queries take a few
seconds over 500 million documents.  Setting $t = \texttt{Entity}$,
the root type, and asking for a stream of all entities in qualifying
snippets, slows down the query by a small factor.  This is all that
our algorithm needs from the corpus; we did not focus on query time
because standard caching techniques and tighter code can improve it
trivially.

\subsection{Queries with ground truth}
\label{sec:Expt:ourGTdata}

We use 709 entity search queries collected from many years
of TREC and INEX competitions, along with relevant and irrelevant
entities.  Two paid masters students, familiar with Web search
engines, read the full TREC/INEX description of entity search queries
and wrote out queries they would naturally issue to a commercial
search engine.  They also selected the best (as per their judgment)
type from YAGO for each query, as ground truth.  This data is
\href{https://docs.google.com/spreadsheet/ccc?key=0AnsqzHjpPcG4dE1haGV4bXRFV01rMG5YYklveF9tQmc#gid=1}{publicly available} at \href{http://bit.ly/WSpxvr}{bit.ly/WSpxvr}.  Launching the queries with the known types resulted in
380 million snippets supporting candidate entities; these are also available on request.
We also performed type prediction (Section~\ref{sec:Disc:TypePredict}) on dataset provided in~\cite{BalogN2012TargetTypeEntityQueries}. Since this dataset does not contain ground truth of relevant entities for each query, we did not test entity ranking.

\subsection{Generic and ``perfect'' baselines}

The ranking accuracy of a reasonable query interpreter algorithm in
our framework will lie between two baselines:
\begin{description}
\item[Generic:] The generic baseline assumes zero knowledge of query
  types, instead using $t = \texttt{Entity}$, the root/s of the type
  hierarchy in the catalog.
\item[``Perfect'':] The ``perfect'' baseline assumes
  complete (human-provided) knowledge of the type and uses it in the
  semistructured query launched over the catalog and annotated corpus.
\end{description}
Of course, even ``perfect'' may perform poorly in some queries,
because of lack of support for relevant entities in the corpus,
snippets incorrectly or not annotated (both false positive and
negative), or incorrect absence of paths between types and entities in
the catalog.  It is also possible for an algorithm (including ours) to
perform worse than generic on some queries, by choosing a particularly
unfortunate type, but obviously it should do better than generic on
average, to be useful.

\subsection{Measurements and results}

As is standard in entity ranking research, we report NDCG at
various ranks, mean reciprocal rank (MRR, not truncated) and mean
average precision (MAP) at the entity (not document) level.  Space
constraints prevent us from defining these; see Liu
\cite{Liu2009LearningToRank} for details.  For Discriminative, 
$C$ is tuned by 5-fold cross validation at the query level.  For Generative,
we swept over $\alpha,\beta,\gamma,\delta$ in powers of 10
(e.g. $10^{-5}, 10^{-4}, \ldots, 1$).

\begin{figure}[h]
\begin{center}
\begin{tabular}{l|r|r|r|r}
& {Generic} & {Generative} &
{Discriminative} & {Perfect} \\
\hline
MAP & $0.323^\downarrow$$^\downarrow$ & $0.414^\downarrow$ & \textbf{0.462} & 0.644\\
\hline
MRR & $0.332^\downarrow$$^\downarrow$ & $0.432^\downarrow$ & \textbf{0.481} & 0.664
\end{tabular}
\end{center}
 \centering\includegraphics[width=20em]{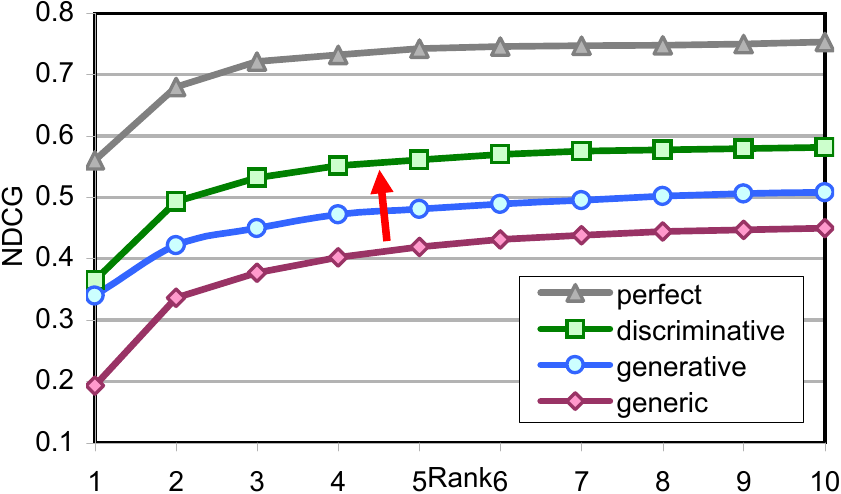}
\caption{Generic, generative, discriminative and ``perfect''
  accuracies.}
  \label{fig:Bracketed}
\end{figure}

\subsubsection{Our algorithms vs.\ generic and perfect}

For our techniques to be useful, they must bridge a substantial
part of the gap between the generic lower bound and the perfect upper
bound.  Figure~\ref{fig:Bracketed} confirms that Generative bridges
28\% of the MAP gap between generic and perfect, whereas
discriminative is significantly better at 43\%.  MRR and NDCG follow
similar trends.  All gaps are statistically significant 
at 95\% confidence level (indicated by $\downarrow$).

\begin{figure}
\centering\includegraphics[width=20em]{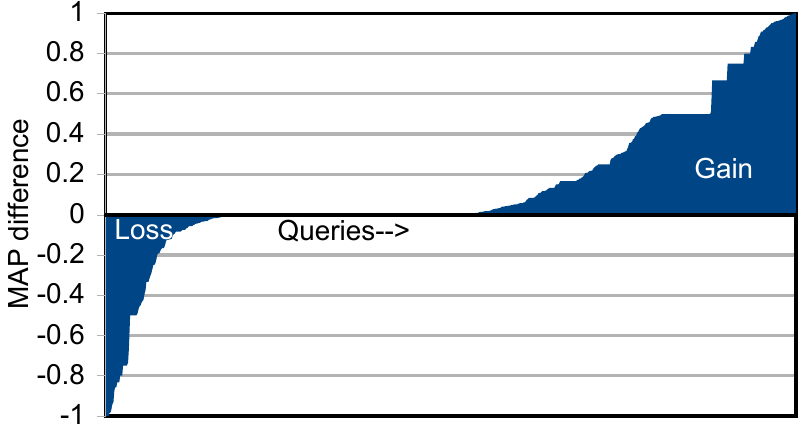}
  \caption{MAP of discriminative minus map of generic, 
compared query-wise between generic and discriminative.
    Below zero means discriminative did worse than generic on that
    query.  Queries in (arbitrary) order of discriminative AP gain.}
  \label{fig:MapByQuery}
\end{figure}

Figure~\ref{fig:Bracketed} is aggregated over all queries.
Figure~\ref{fig:MapByQuery} focuses on average precision disaggregated
into queries, comparing discriminative against generic.
While some queries are damaged by discriminative, many more are improved.  

Failure analysis revealed residual $(t,\vec z)$ ambiguity, coupled
with lack of $\in^+$ or $\subseteq^+$ paths in an incomplete catalog
to be the major reasons for losses on some queries.
Even though there is some ground yet to cover to reach ``perfect'' levels,
these results show there is much hope for automatically interpreting
even telegraphic queries.

\subsubsection{Benefits of annealing optimization}

Figure~\ref{fig:AnnealBenefits} shows that discriminative with our
entropy-based annealing protocol performs significantly (marked 
with ``$\downarrow$'') better than the scheme proposed by Bergeron \etal
\cite{BergeronZBB2008MIR}.  This may be of independent interest in
multiple instance ranking and max-margin learning with latent
variables.

\begin{figure}[h]
\centering
\begin{tabular}{|r|r|r|}
\hline
\, & Bergeron \eqref{eq:PreAnneal} & Entropy \eqref{eq:Anneal} \\
\hline
MAP & $0.416^\downarrow$ & \textbf{0.462} \\
\hline
MRR & $0.432^\downarrow$ & \textbf{0.481} \\
\hline
\end{tabular}
  \caption{Benefits of annealing protocol.}
  \label{fig:AnnealBenefits}
\end{figure}

\subsubsection{Benefits of joint inference}
\label{sec:Expt:JointBenefit}

A central premise of our work is that joint inference is better than a
two-stage process (predict type, launch query).
To test the essence of this hypothesis, we
run our system, throw away the ranked entity list,
and only retain the predicted type
(Section~\ref{sec:Disc:TypePredict}), then launch a query restricted
to this type (Section~\ref{sec:Expt:QueryProcessor}) and measure entity
ranking accuracy.  

\begin{figure}[h]
\begin{center}
\begin{tabular}{l|r|r|r|r}
& {Joint} & {2-stage} &
{2-stage} & {2-stage} \\
& &  ($k=1$) &  ($k=5$) & ($k=10$) \\
\hline
MAP & \textbf{0.462} & $0.370^\downarrow$ & $0.361^\downarrow$ & $0.365^\downarrow$\\
\hline
MRR & \textbf{0.481} & $0.384^\downarrow$ & $0.375^\downarrow$ & $0.377^\downarrow$\\  
\end{tabular}
\end{center}
\centering\includegraphics[width=20em]{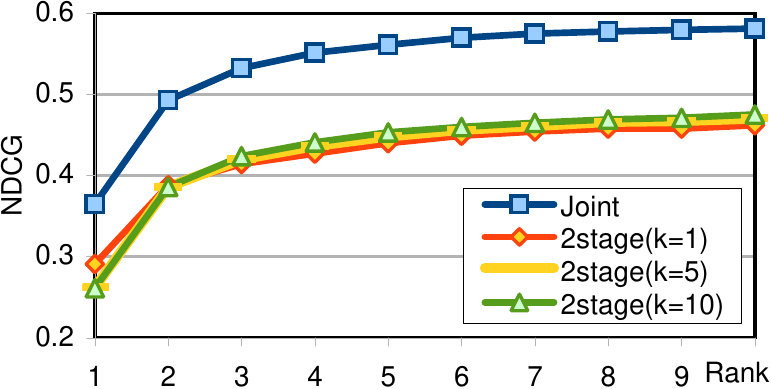}
\caption{Joint inference improves entity ranking quality
compared to 2-stage.}
    \label{fig:JointVsStaged}
\end{figure}

Figure~\ref{fig:JointVsStaged} shows that the result is significantly
(shown by ``$\downarrow$'') less accurate than via joint inference,
even after tuning $k$, which indicates that no \emph{single} inferred type
may retain enough information for the best entity ranking.

\subsubsection{Comparison with B\&N's type prediction}

B\&N \cite{BalogN2012TargetTypeEntityQueries} proposed two models, of
which the ``entity-centric'' model was generally superior.  Each
entity $e$ was associated with a textual description (e.g., Wikipedia
page) which induced a smoothed language model $\theta_e$.  B\&N
estimate the score of type $t$ as
\begin{align}
  \Pr(q|t) &= \sum_{e \in^+ t} \Pr(q|\theta_e) \Pr(e|t),
\end{align}
where $\Pr(e|t)$ was set to uniform.  Note that no corpus (apart from
the one of entity descriptions) was used.  The output of B\&N's
algorithm (hereafter, ``B\&N'') is a ranked list of types, not
entities.  We implemented B\&N, and obtained accuracy closely matching
their published numbers, using the DBpedia catalog with 358 types,
and 258 queries (different from our main query set and testbed).

\begin{figure}[h]
\centering
\begin{tabular}{|r|r|r|r|r|}
\hline
~ & B\&N & Discr($k=1$) & Discr($k=5$) & Discr($k=10$)\\
\hline
MAP & 0.33 & 0.33 & 0.384 & 0.390\\
\hline
\end{tabular}
\centering\includegraphics[width=20em]{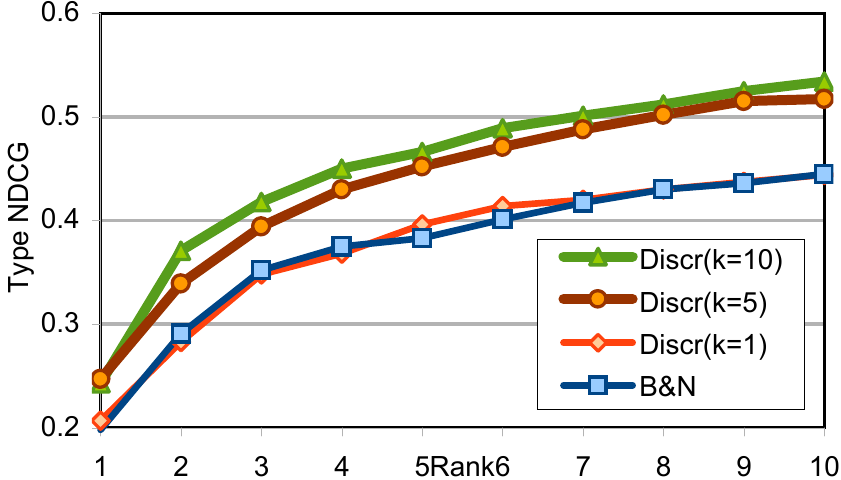}
  \caption{Type prediction by B\&N vs.\ discriminative.}
  \label{fig:BnType}
\end{figure}

We turned our system into a type predictor
(Section~\ref{sec:Disc:TypePredict}), and also used DBpedia like B\&N and 
compared type prediction accuracy on dataset provided in \cite{BalogN2012TargetTypeEntityQueries}. Results are shown in Figure~\ref{fig:BnType}.  At $k=1$, our
discriminative type prediction matches B\&N, and larger $k$ performs
better, owing to stabilizing consensus from lower-ranked entities.  
Coupled with the results in
Section~\ref{sec:Expt:JointBenefit}, this is strong evidence that
our unified formulation is superior, even if the goal is type prediction.

\subsubsection{Comparison with B\&N-based entity ranking}

A type prediction may be less than ideal, and yet entity prediction
may be fine.  One can take the top type
predicted by B\&N, and launch a query (see
Section~\ref{sec:Expt:QueryProcessor}) with that type restriction.  To
improve recall, we can also take the union of the top $k$ predicted
types.  The search results in a ranked list of entities, on which we
can compute entity-level MAP, MRR, NDCG, as usual.  In this setting,
both B\&N and our algorithm (discriminative) used YAGO as the catalog.
Results for our dataset (Section~\ref{sec:Expt:ourGTdata}) are shown in Figure~\ref{fig:BNEnt}.

\begin{figure}[h]
\centering
\begin{tabular}{r|r|r||r|r}
$k$ & MAP & MRR & $\%\mathcal{Q}$ better & $\%\mathcal{Q}$ worse \\
\hline 
1 & 0.066 & 0.068 & 5.50 & 88.58\\
5 & 0.137 & 0.144 & 15.80 & 76.30\\
10 & 0.171 & 0.180 & 20.73 & 69.53 \\
15 & 0.201 & 0.211 & 24.54 & 63.47 \\
20 & 0.204 & 0.215 & 26.80 & 60.51 \\
25 & 0.222 & 0.233 & 29.34 & 56.84 \\
30 & 0.232 & 0.244 & 29.76 & 55.01 \\
\hline 
Generic & 0.323 & 0.432 & --- & --- \\
\end{tabular}
  \caption{B\&N-driven entity ranking accuracy.}
  \label{fig:BNEnt}
\end{figure}

We were surprised to see the low entity ranking accuracy (which is why
we recreated very closely their reported 
type ranking accuracy on DBpedia).  Closer
scrutiny revealed that the main reason for lower accuracy was changing
the type catalog from DBpedia (358 types) to YAGO (over 200,000
types).  Entity ranking accuracy is low because B\&N's type prediction
accuracy is very low on YAGO: 0.04 MRR, 0.04 MAP, and 0.058 NDCG@10.  For
comparison, our type prediction accuracy is
0.348 MRR, 0.348 MAP, and 0.475 NDCG@10.
This is entirely because of corpus/snippet signal:
if we switch off snippet-based features $\phi_4$, our accuracy 
also plummets.  The moral seems to be, large organic type catalogs
provide enough partial and spurious matches for \emph{any} choice of hints, so
it is essential (and rewarding) to exploit corpus signals.

\begin{figure}[h]
\centering\includegraphics[width=20em]{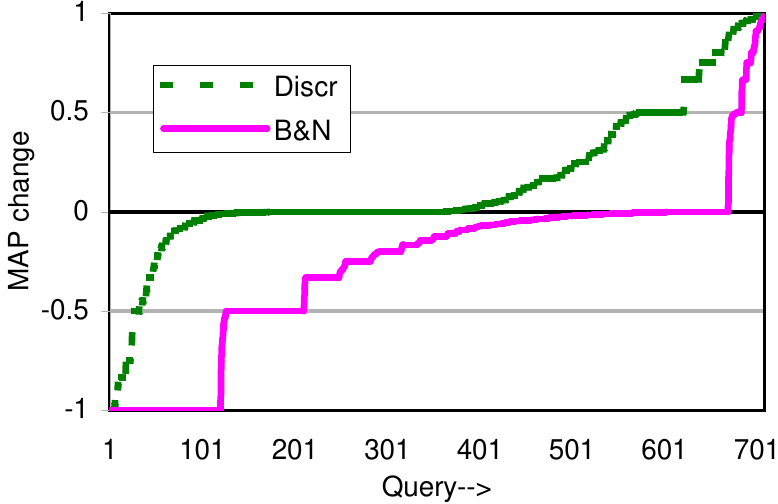}
  \caption{2-stage entity ranking via B\&N does boost accuracy
for some queries, but the overall effect is negative.  Joint
interpretation and ranking also damages some queries but 
improves many more.}
  \label{fig:BnEntApVsGeneric}
\end{figure}

On an average, B\&N type prediction, followed by query launch, seems
worse than generic.  This is almost entirely because of choosing bad
types for many, but not all queries.  There \emph{are} queries where
B\&N shows a (e.g., MAP) lift beyond generic, but they are just too
few (Figure~\ref{fig:BnEntApVsGeneric}).

\begin{figure}[h]
\centering
\begin{tabular}{r|r|r}
$k$ & MAP & MRR \\
\hline 
1 & 0.135 & 0.145\\
5 & 0.240 & 0.250\\
10 & 0.295 & 0.305\\
\hline 
Discr & 0.422 & 0.437\\
\end{tabular}
\centering\includegraphics{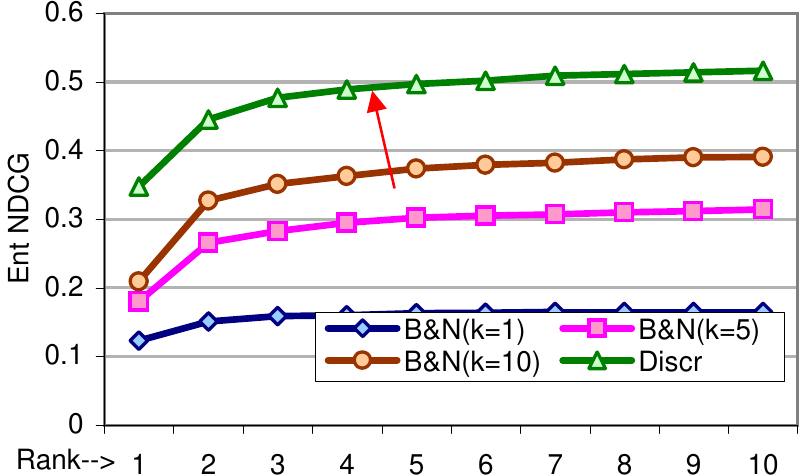}
  \caption{Entity ranking accuracy using DBpedia types.}
  \label{fig:DBpediaEnt}
\end{figure}

\subsubsection{Coarse DBpedia types with Web corpus}
\label{sec:Expt:DBpediaWeb}

A plausible counter-argument to the above experiments is that, by
moving from only 358 DBpedia types to over 20,000 YAGO types, we are
making the type prediction problem hopelessly difficult for B\&N, and
that this level of type refinement is unnecessary for high accuracy in
entity search.  We modified our system to use types from DBpedia, and
correspondingly re-indexed our Web corpus annotations using DBpedia
types.  As partial confirmation of the above hypothesis, the entity
ranking accuracy using B\&N did increase substantially.  However, as
shown in Figure~\ref{fig:DBpediaEnt}, the entity ranking accuracy
achieved by our discriminative algorithm remains unbeaten.  Also
compare with Figure~\ref{fig:Bracketed} --- whereas B\&N improves by
coarsening the type system, our discriminative algorithm seems to be
degraded by this move.

\subsubsection{DEANNA on telegraphic queries}
\label{sec:Expt:Deanna}

We also tried to use the \href{https://d5gate.ag5.mpi-sb.mpg.de/deannaWeb/deannaIlpNew.htm}{Web interface} to send a sample of our
telegraphic queries and their well-formed sentence counterparts to
DEANNA \cite{YahyaBERTW2012DEANNA} and receive back the
interpretation.  We manually inspected their output.  Some anecdotes
are shown in Figure~\ref{fig:DEANNA}.  The queries are from
Figure~\ref{fig:Telegraph}.  None of the telegraphic queries was
successfully interpreted.  The well-formed questions saw partial
success.

\begin{figure}[h]
\centering
\begin{tabular}{l|p{.45\hsize}|p{.4\hsize}}
QID & Well-formed & Telegraphic \\
\hline \hline
Q1 & Missing target type & Empty
 \\ \hline 
\raggedright
Q2 & Incorrect, missed Wikipedia
type ``list of cities in China'' & Incorrect fragments
 \\ \hline
Q3 & Incorrect target type (painting) & Empty
 \\ \hline
Q4 & Incorrect fragments & Incorrect fragments
 \\ \hline
Q5 & Incorrect fragments & Empty
 \\ \hline
Q6 & No target type & Empty
\end{tabular}
  \caption{DEANNA interpretations of some of our queries.}
  \label{fig:DEANNA}
\end{figure}


\section{Conclusion}
\label{sec:End}

We initiated a study of generative and discriminative formulations for
joint query interpretation and response ranking, in the context of
targeted-type entity search needs expressed in a natural
``telegraphic'' Web query style.  Using 380 million snippets from
a Web-scale corpus with 500
million documents annotated at 8 billion places
with over 1.5 million entities and 200,000
types from YAGO, We showed experimentally that jointly interpreting
target type and ranking responses
is superior to a two-phase interpret-then-execute paradigm.

Our work opens up several directions for further research.  Our notion
of selectors can be readily generalized to allow mentions of entities
as literals \cite{GuoXCL2009nerq, PantelLG2012EntityTypeIntent} in the
query.  More sophisticated training using
bundle methods may further improve the discriminative formulation.
Finally, modeling listwise \cite{Liu2009LearningToRank} losses may 
also help.



\begin{small}
\bibliographystyle{abbrv}
\bibliography{voila}  
\end{small}

\end{document}